\newcommand{\be}{\begin{eqnarray}}
\newcommand{\ee}{\end{eqnarray}}
\newcommand{\nn}{\nonumber}

\documentclass[12pt]{article}

\usepackage{amsmath,amssymb}
\usepackage{graphicx}
\usepackage{cite}
\usepackage{bm}
\begin{document}

\renewcommand{\thefootnote}{\fnsymbol{footnote}}

\vskip 15mm

\begin{center}

{\Large Cryptogauge symmetry and cryptoghosts
 for crypto-Hermitian Hamiltonians}

\vskip 4ex

A.V. \textsc{Smilga}\,$^{1}$,

\vskip 3ex

$^{1}\,$\textit{SUBATECH, Universit\'e de
Nantes,  4 rue Alfred Kastler, BP 20722, Nantes  44307, France
\footnote{On leave of absence from ITEP, Moscow, Russia.}}
\\
\texttt{smilga@subatech.in2p3.fr}
\end{center}

\vskip 5ex

\begin{abstract}
\noindent We discuss the Hamiltonian $H = p^2/2 - (ix)^{2n+1}$ and the mixed Hamiltonian 
$H_{\rm mixed} = (p^2 + x^2)/2 - g(ix)^{2n+1}$.
The Hamiltonians $H$ and in some cases also $H_{\rm mixed}$ are 
crypto-Hermitian in a sense that, in spite of their apparent non-Hermiticity, a quantum spectral problem can be formulated 
such that the spectrum is real. We note  that the corresponding classical Hamiltonian system can be 
treated as a gauge system, with imaginary part of the Hamiltonian playing the role of the first class constraint. 

 Several different nontrivial quantum problems can be formulated on the basis of this classical problem. 
We formulate and solve some such problems. We consider then the mixed Hamiltonian and find that its spectrum 
 undergoes in certain cases a rather amazing transformation when the coupling $g$ 
is sent to zero. There is an infinite set of exceptional points $g_\star^{(j)}$ where a couple of eigenstates of $H$ 
coalesce and their eigenvalues cease to be real.

 When quantization is done in 
the most natural way such that gauge constraints are imposed on quantum states, the spectrum should
 not be positive definite,
but must involve the negative energy states ( ghosts).  
 We speculate that, in spite of the appearance of ghost states, unitarity 
might 
still be preserved. 

\end{abstract}

\renewcommand{\thefootnote}{\arabic{footnote}}
\setcounter{footnote}0
\setcounter{page}{1}

\section{Introduction}
For certain apparently complex Hamiltonians, the spectral problem
can be formulated such that the spectrum has a perfectly ``normal'' form with bounded from below real energies. 
Such Hamiltonians can thus be called ``crypto--Hermitian'' or ``cryptoreal''.
Apparently,  such 
crypto-Hermitian Hamiltonians were first discussed in association with Reggeon field theory back in 1976 \cite{Regge}.
Somewhat later, crypto-Hermitian Hamiltonians were  considered by mathematicians in a more habitual Schr\"odinger setup.
 Gasymov observed that the Schr\"odinger operator with 
 certain complex periodic potentials, like $V(x) = e^{ix}$, has a real spectrum \cite{DAN}. 
In Ref. \cite{ital}, it was 
proved that the spectrum of the Hamiltonian with complex potential $V(x) = x^2 + i\beta x^3$ is real for small enough
$\beta$.  General properties of crypto-Hermitian (or {\it quasi-Hermitian} as the authors called this property) 
operators were studied in Ref.\cite{Geyer}.

Before going further, a comment on the terminology is in order. 
At the moment, there is no unique generally adopted name for  Hamiltonians of this kind. Besides  {\it quasi-Hermitian}, the
term {\it pseudo-Hermitian} is also often used. A Hamiltonian is usually called pseudo-Hermitian if it satisfies the property
 \be
\label{eta} 
H^\dagger = \eta H \eta^{-1}
 \ee
 with some Hermitian 
invertible $\eta$. However, this does not guarantee yet that the spectrum is real. 
To this end, the operator $\eta$ should be representable as \cite{Most1} 
 \be
\label{etaOO}
\eta = O^\dagger O
 \ee
  or, equivalently, the norm
$\langle \psi \,| \eta  \psi \rangle$ should be positive definite for any nonzero Hilbert space vector 
$\psi$ \ \cite{Pease},\cite{Geyer}.
Anyway, the semantics of  the words {\it quasi-Hermitian} or {\it pseudo-Hermitian} is ``not quite Hermitian'' 
with a flavour of inferiority, ``second-rankness'' compared to {\it Hermitian}.
 For example, pions are pseudo-Goldstone particles
meaning that they are {\it not} Goldstone particles. But we want to emphasize that, if the spectrum of the Hamiltonian is real, 
the latter almost always
\footnote{``Almost'' means away from  exceptional points \cite{Heiss} where the Hamiltonian involves Jordan blocks. 
We will discuss this issue later.}
 {\it is} in fact Hermitian when looking at it through proper glasses, i.e. when defining the norm in  Hilbert
space in a proper way. It was proved in \cite{Most1} that the Hamiltonian with real non-degenerate spectrum must satisfy 
the properties (\ref{eta}, \ref{etaOO}). Then $\eta$ defines the norm with respect to which the Hamiltonian $H$ is Hermitian,
 while the Hamiltonian $\tilde H = OHO^{-1}$ is manifestly Hermitian with respect to the standard norm. In other words, the
characterization ``crypto-Hermitian'' 
\footnote{It was used first in Ref. \cite{Feinberg} }
(Hermitian in disguise) reflects more adequately, in our opinion,  the essence of the phenomenon, 
and we will stick to it in this paper.
\footnote{Let us repeat for clarity: our {\it crypto-Hermiticity} means exactly the same as 
 {\it quasi-Hermiticity} of Ref.\cite{Geyer} (a quasi-Hermitian Hamiltonian was defined there as the Hamiltonian that is Hermitian with respect to
a generalized positive definite  norm $\langle \psi| \eta \psi \rangle$,\ $\eta^\dagger = \eta$), the same  as {\it $\eta$-Hermiticity} as defined
in  Ref.\cite{Pease}
and the same as {\it pseudo-Hermiticity} (\ref{eta}) with additional requirement (\ref{etaOO}).}

The modern 
history begins with the beautiful paper \cite{BenBoe} (see also the recent review \cite{Bender}), 
where this property was observed for a wide class of 
$PT$-symmetric polynomial potentials, like $V(x) = ix^3$. 
\footnote{ A $PT$-symmetric potential $V(x)$ enjoys the property $V^*(-x) = V(x)$.}. It was found to be discrete and real.

Since then, many crypto-Hermitian Hamiltonians 
have been discovered. We can mention the paper \cite{Jia} where the spectrum of the Hamiltonian with  hyperbolic and 
generalized hyperbolic
complex PT-symmetric potentials was shown to be  real in many cases.
The simplest example of such cryptoreal hyperbolic problem is the problem with the potential
 \be
\label{Jiahyp}
 V(x) \ =\ - \frac {V_1}{\cosh^2 x} + \frac {iV_2 \sinh x}{\cosh^2 x} 
 \ee
with $V_1 > 0$ and $|V_2| < V_1 + 1/4$.

In recent \cite{IvSmi}, it was shown that apparently complex Hamiltonians obtained after
so called {\it nonanticommutative} deformations \cite{Seiberg} of certain supersymmetric quantum-mechanical and 
field theory models 
are in fact crypto-Hermitian and enjoy a real spectrum.

The problems with the potential $V(x) = e^{ix}$ or the potential (\ref{Jiahyp}) admit
explicit analytic solutions.
 In  \cite{BenBoe}, reality of the spectrum for the potentials $V(x) = x^2 (ix)^\epsilon $ , $\epsilon \geq 0$, 
was demonstrated explicitly by numerical solution of the corresponding Schr\"odinger equations supplemented by 
semiclassical analysis. 
Later, a rigorous proof for the discreteness and reality of the spectrum in this problem was constructed \cite{Dorey}.
 In Ref.\cite{Most}, it was shown that the Hamiltonians like 
  \be
 \label{2n+1}
H \ =\ \frac {p^2 + x^2}2 - g(ix)^{2n+1}
 \ee
can be represented  for small $g$  in the form (\ref{eta}, \ref{etaOO}). In other words, they
can be obtained by a non-unitary transformation, $H = e^{-R} \tilde H e^{R}$, out of  a manifestly Hermitian Hamiltonian
$\tilde H$. The Hamiltonian $\tilde H$ and the operator $R \equiv \ln O $ are calculated perturbatively as an infinite series 
in the coupling constant $g$.

In this paper we suggest an approach capitalizing on a certain hidden gauge symmetry characteristic 
of crypto-Hermitian systems. The origin of this symmetry is very simple \cite{Koshi1}-\cite{Koshi3}. Consider a system with 
one dynamical degree of freedom. The classical Hamiltonian is a function ${\cal H}(p,x)$, which 
may be real or complex. Let us complexify the phase space variables,
 \be
\label{compl}
x \to z = x + iy,\ \ \ \ \ \ \ \ \ \ \ p \to \pi = p -iq \ ,\nn \\
 {\cal H}(p,x) \to {\cal H}(\pi,z) \ =\ H(p,q ;x,y) + iG(p,q ;x,y) \ ,
 \ee
where $H$ and $G$ are real functions satisfying the Cauchy-Riemann relations
 \be
\label{kosh}
\frac {\partial H}{\partial p} + \frac {\partial G}{\partial q} 
= \frac {\partial H}{\partial y} + \frac {\partial G}{\partial x} = \frac {\partial H}{\partial q} - 
\frac {\partial G}{\partial p} = \frac {\partial H}{\partial x} - \frac {\partial G}{\partial y} = 0\ .
 \ee
Two important properties follow :
  \begin{itemize}
  \item The function $ H(p,q ;x,y)$ can be treated as the Hamiltonian of a new system with double set 
of degrees of freedom. Indeed, the real and imaginary parts of the complexified equations of motion for 
the original system, 
$$ \dot \pi = -\partial {\cal H}/\partial z,\ \ \ \ \ \dot z = \partial {\cal H}/\partial \pi \ ,$$
coincide in virtue of (\ref{kosh}) with the Hamilton equations of motion derived from $ H(p,q ;x,y)$. 
 \item The Poisson bracket
 $$ \{H,G \}_{P.B.} = \frac {\partial H}{\partial x} \frac {\partial G}{\partial p} + 
\frac {\partial H}{\partial y} \frac {\partial G}{\partial q} - \frac {\partial H}{\partial p} 
\frac {\partial G}{\partial x} - \frac {\partial H}{\partial q} \frac {\partial G}{\partial y} $$
vanishes. This means that $G$ is an integral of motion for the system described by  $H$. The space 
of all classical trajectories is thus divided into classes characterized by a definite value of $G$. 
The class with $G=0$ represents a particular interest. The condition $G=0$ can be interpreted as a 
first class constraint and the dynamical system with the Hamiltonian $H$ supplemented by the 
constraint $G=0$ is a {\it gauge} system. 
 \end{itemize}

The plan of the paper is the following. In the next section, we consider from this angle the
 simplest possible problem --- the complexified oscillator. We note that this classical problem
 has at least three different quantum counterparts: 
\begin{enumerate}
\item One can impose the analyticity constraint on the wave function,
$\partial \psi/\partial \bar z = 0$ and solve the Schr\"odinger equation in the vicinity of the 
real axis. In this case, we reproduce the standard oscillator spectrum $E_n = 1/2 + n$.
The same spectral problem is obtained when the gauge constraint is resolved at the classical level
with the gauge choice $y=0$.
\item One can impose the analyticity constraint and solve the Schr\"odinger equation in the 
vicinity of the imaginary axis. In this case, the spectrum 
$E_n = -1/2 - n$ involves negative energies and is bounded from above rather from below. 
 The same spectral problem is obtained when the gauge constraint is resolved at the classical level
with the gauge choice $x=0$.
 \item Finally, one may not require analyticity, but rather impose, following Dirac, the gauge constraint 
$\hat G \Psi = 0$ on quantum states. In this case, the spectrum is $E_n = n$, where $n$ can be positive, 
zero, or negative. Still, the quantum problem is well defined, and the evolution operator is unitary.
 \end{enumerate}

In Sect. 3, we consider the classical dynamics of the Hamiltonian
  \be
  \label{bezosc}
H = \frac{\pi^2}2 - (iz)^{2n+1}\ .
  \ee 
We find different sets of trajectories with positive, and also with {\it negative} energies. 
Sect. 4 is devoted to the quantum dynamics of (\ref{bezosc}) and of the mixed Hamiltonian
 (\ref{2n+1}), 
with the analyticity constraint 
imposed on wave functions. The complex plane of $z$ is divided then into several regions. In
some of them, the spectrum is discrete, in some others - continuous or 
empty.  For $n=1$, we reproduce the  results of Ref.\cite{BenBoe}. 
For $n > 1$, one can formulate $n$ {\it different} spectral problems 
with discrete positive definite
spectrum formulated in the different regions of the complex plane.
\footnote{That was observed in Ref.\cite{BeBoMe}. The proof of reality and discreteness of the spectrum for 
all {\it symmetric} (see below) spectral problems was given in Ref.\cite{Shin}.} 
  When $g \to 0$, the spectrum of the mixed 
Hamiltonian approaches the oscillator spectrum, but for the problems formulated in the sectors not comprising 
 real axis, the transformation pattern is very nontrivial involving an infinite  set
of ``phase transitions'' in the coupling. At each of such ``phase transition'' (or exceptional \cite{Heiss} point $g_*^{(j)}$), 
a pair of eigenstates of the mixed Hamiltonian coalesce such that at this very point the Hamiltonian 
involves a nondiagonalizable Jordan block. Beyond this point ($ g < g_*^{(j)}$), a pair of complex conjugate 
eigenvalues should appear. 
In other words, the Hamiltonian (\ref{2n+1})  is cryptoreal in this situation only for large enough $g$ 
(and, obviously, it is Hermitian for $g = 0$).
 The last section is devoted as usual to discussions. In particular, we discuss the Dirac 
spectral problem for the Hamiltonian (\ref{bezosc}) when the gauge constraint  
is imposed  on the wave functions as an operator condition. 
 This problem has no analytic solution and is difficult to resolve numerically. Still, 
based on semiclassical reasoning, we argue that, similar to what we had
 in the oscillator case, the spectrum there might be discrete and unbounded both from below and 
  above.  We also point out the 
similarity of this problem to some other previously analyzed by 
us systems, which are described by higher derivative Lagrangians and involve ghosts. 
We speculate that, in spite of their presence, unitarity is not violated.

   \section{Complex oscillator.}
Consider the complex Hamiltonian
  \be
  {\cal H}(\pi, z) =\ \frac {\pi^2 + z^2}2 \ .
  \ee
Its real and imaginary parts are 
  \be
 \label{Hosc}
H \ =\ \frac {p^2 + x^2}2 - \frac {q^2 + y^2}2
 \ee
and
 \be
\label{Gosc}
G \ =\ -pq + xy \ .
 \ee
Consider the classical dynamics of $H$. The classical trajectories are
 \be
\label{trajH}
x = A \sin(t+\phi_1),\ \ p = A\cos(t + \phi_1),\ \ \ y = B\sin(t + \phi_2),\ \ \ q = -B\cos(t + \phi_2)\ .
\ee
Generically, they have complex energies. If we require the energies to be real, i.e. impose the constraint 
$G = 0$, the relation
 \be
\label{ABfi}
AB\cos(\phi_1 - \phi_2) \ =\ 0 
  \ee
follows. For each value of the energy, positive or negative, there is a set of trajectories (cofocal ellipses) 
with the same period (see Fig. 1). 

\begin{figure}[h]
   \begin{center}
 \includegraphics[width=4.0in]{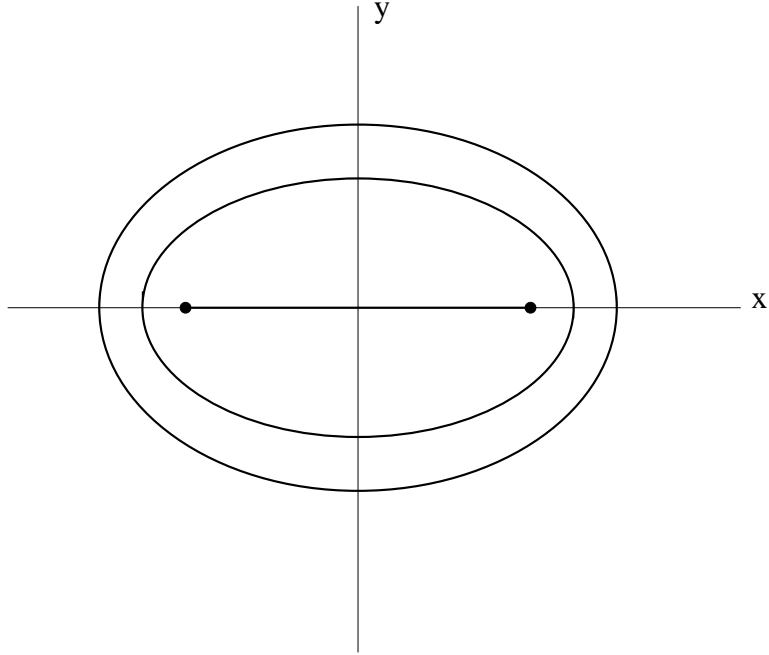}
    \end{center}
\caption{Family of oscillator trajectories with the same positive energy. For negative energies, 
the picture is rotated by $\pi/2$.}
\label{ellipsy}
\end{figure}

In the case under consideration, the period is the same for {\it all} energies, but this is the specifics
of oscillator. The fact that the period is the same for all trajectories of a {\it given} energy has, 
however, a general nature.
In fact, it is a consequence of the gauge symmetry of the problem.

The latter is simply the symmetry generated by the constraint $G$. Infinitesimally \cite{Koshi3}, 
\be
\label{gaugebezlam}
\delta_G x &=& -\alpha \{G, x \}_{P.B.} = - \alpha q,\ \ \ \ \ \delta_G y = \ \ \ -\alpha \{G, y \}_{P.B.}
-\alpha p,\nn \\ 
\delta_G p &=& -\alpha \{G, p \}_{P.B.} =  -\alpha y, \ \ \ \ \ \delta_G q = \ \ \ 
 -\alpha \{G, q \}_{P.B.} = -\alpha x \ .
 \ee
This is a phase space symmetry. To represent it as a conventional gauge symmetry acting only on the 
coordinates, one should
introduce the Lagrange multiplier $\lambda(t)$ and write the canonical Lagrangian as
 \be
\label{Lcan}
L \ =\  p {\dot x} + q {\dot y} - H - \lambda G \ ,
 \ee
  Expressing out the momenta,
\be
\label{momenta}
p \ =\ \frac{{\dot x} - \lambda {\dot y}}{1 + \lambda^2}\ ,\ \ \ \ \ \ 
q \ =\ -\frac{{\dot y} + \lambda {\dot x}}{1 + \lambda^2}\ ,
  \ee
we obtain
 \be
\label{Losc}
L = \frac {{\dot x}^2 - {\dot y}^2 - 2\lambda {\dot x}{\dot y}}{2(1 + \lambda^2)} + \frac {y^2-x^2}2
- \lambda xy \ .
 \ee
The gauge transformations amount to shifting the Lagrange multiplier $\lambda$ by (a derivative of) 
an arbitrary function of time $\dot \alpha (t)$,  
 supplemented by the transformations of dynamic variables $x,y$ generated
by the constraint $G$. 
 \be
\label{gauge}
 \delta_{\rm gauge} x &=& -\alpha q = \frac {\alpha({\dot y} + \lambda {\dot x})}{1 + \lambda^2}\ , \nn \\  
\delta_{\rm gauge} y &=& -\alpha p = -\frac {\alpha({\dot x} - \lambda {\dot y})}{1 + \lambda^2}\ , \nn \\
 \delta_{\rm gauge} \lambda &=& {\dot \alpha}\ .
  \ee
Indeed, one can explicitly verify that the Lagrangian (\ref{Losc}) is invariant, up to a total derivative, 
with respect to the transformations (\ref{gauge}). 

The transformations $\delta x$ and $\delta y $ 
in Eqs.(\ref{gaugebezlam}, \ref{gauge}) have a clear meaning. Any Hamiltonian system is invariant with respect 
to time 
translations $t \to t - a$ that transform a solution $z(t) \to z(t-a)$. Their generator is the Hamiltonian
$H$. In our case, however, besides $H \equiv {\rm Re}( {\cal H})$, we have another integral of motion
$G \equiv {\rm Im} ({\cal H})$. It generates a shift of time by an {\it imaginary} amount, $t \to
t - i\alpha$ and transforms $z(t) \to z(t - i\alpha)$. Infinitesimally, this coincides with Eq.(\ref{gauge}) 
(with partial
gauge fixing $\lambda = 0$).

The shift $z(t) \to z(t-a)$ is the shift along the trajectory, leaving it unchanged. 
But the shift $z(t) \to z(t -i\alpha)$ transforms one trajectory into another. It  is this shift which
 relates different ellipses in Fig. 1 
[it is straightforward to check by substituting for $z(t)$ the exact analytic solution (\ref{trajH}) with
$\phi_1 - \phi_2 = \pi/2$]. Such families of closed trajectories of a given energy and the same period
(obviously, if $z(t)$ is periodic, $z(t - i\alpha)$ is also periodic with the same real period) exist also 
for more complicated cases. We will discuss it in the next section.

 Let us go over to quantum dynamics. There are two basic ways to quantize  gauge systems 
\footnote{This problem was first posed and resolved by Dirac and is treated pedagogically in many books. See e.g.
Ref.\cite{Slavnov}}: 
{\it (i)} by explicitly resolving the constraints and quantizing the Hamiltonian with a reduced number of degrees of freedom;
{\it (ii)} by not resolving the constraints classically, but rather solving the system 
 \be
\label{Schrgauge}
 \hat H \Psi = E \Psi, \ \ \ \ \ \ \ \ \ \hat G \Psi = 0\ .
 \ee
We will see that,  in the case under
consideration, these two approaches are not quite equivalent, in contrast to what is usually assumed !

\begin{itemize}
\item
Let us first try to resolve the constraint $G=0$ classically. This can be done  by fixing the gauge, i.e. by  
imposing the additional constraint $\chi(p,q; x,y) = 0$, where $\{G, \chi \}_{P.B.} \neq 0$ (so that
 the primary constraint
$G = 0$ and the gauge fixing constraint $\chi = 0$ are independent). Resolving the system $G = \chi = 0$,
 we are left with a reduced number
of dynamical variables. Generically, their number is equal  to the number of initial degrees of freedom minus
 the number of primary constraints. 
In our case, $N_{\rm reduced} = 2-1 =1$. One can, for example, choose $\chi = y = 0$. The reduced 
Hamiltonian system will in this case be just
$H^* = (p^2 + x^2)/2$ with the spectrum $E_n = 1/2 + n$. On the other hand, if choosing the gauge 
$\chi = x = 0$, the reduced Hamiltonian
is $H^* = -(q^2 + y^2)/2$ with a  different spectrum $E_n = -1/2 - n$.
In other words, there are two essentially {\it different} gauge choices leading to {\it different} 
reduced Hamiltonians. One can obtain either oscillator 
with positive energies, or oscillator with negative energies, but not both. 

To understand what happened, look again at the trajectories in Fig. 1. They represent, as we have seen, 
gauge copies of one another. 
The gauge fixing procedure should pick out one of
 these copies, while getting rid of all others. And, indeed, the condition $y=0$ does this job by pinpointing 
the trajectory going along the real
axis.  However, none of 
these trajectories are compatible with the condition $x=0$. On the other hand, for the family of the 
trajectories with negative energies, one can 
impose $x=0$ (and pinpoint the trajectory going along the imaginary axis), but not $y=0$.
\footnote{The trajectories in Fig. 1 are related by gauge transformations with constant $\alpha$. But one can easily prove
that one cannot obtain a configuration with $x(t) = 0$ out of a configuration with $y(t) = 0$ by a 
generic gauge transformation (\ref{gauge}). Indeed, the energy functional is positive definite  when $y(t) = 0$ and negative
definite when  $x(t) = 0$.}

Two  spectral problems with positive and negative energies can alternatively be defined using the approach 
of Ref.\cite{BenBoe}. To this end, 
one should require that the wave function represents an analytic function of $z = x+iy$. 
The spectrum $E_n = 1/2+n$ is then realized 
by the standard oscillator functions
continued analytically to complex arguments. For example, the wave function of the ground state is 
$\exp(-z^2/2)$. It falls down exponentially
 on the real axis and also on the lines $z = us,\ \ s \in (-\infty, \infty), \ |{\rm Arg}(u)| < \pi/4$.
The spectrum $E_n = -1/2 - n$ is realized by the functions like $\exp(z^2/2)$ that fall down exponentially along the 
imaginary axis and in the sector
$|{\rm Arg}(z)| > \pi/4$ (see Fig. 2).

\begin{figure}[h]
   \begin{center}
 \includegraphics[width=4.0in]{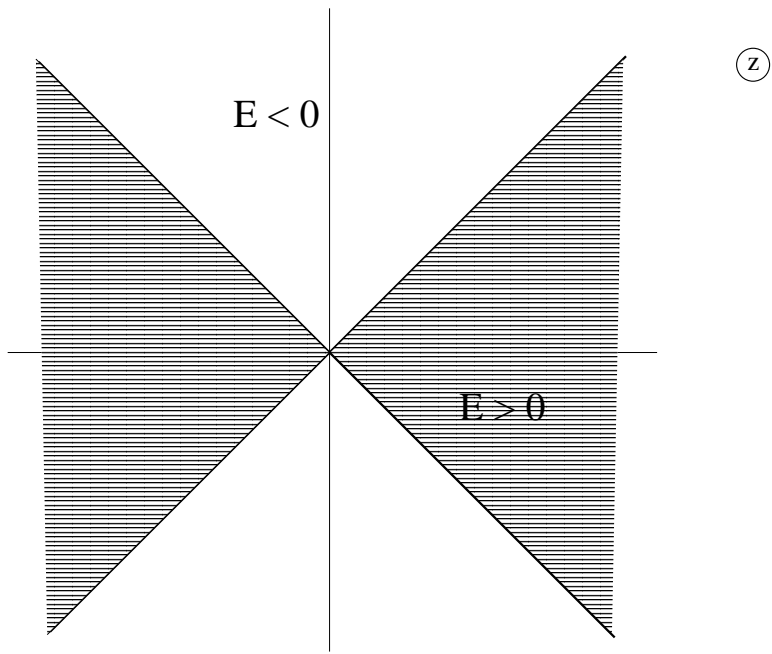}
    \end{center}
\caption{Sectors with positive and negative energies in the spectrum}
\label{sektora}
\end{figure}

The lines Arg$(z) = \pm \pi/4$ are closely related with the {\it Stokes lines} of the oscillator Schr\"odinger equation.
The Stokes lines are defined \cite{Stokes}  as the lines that pass through 
turning points and satisfy the condition
 \be
\label{Stokes}
{\rm Im} \, \left( \int_{z_0}^z \pi(w) dw \right) = \left( \int_{z_0}^z \sqrt{2(E - w^2)}  dw \right)\ =\ 0\ ,
  \ee
( $z_0$ is the position of the turning point). The asymptotes of Stokes lines 
at large values of $|z|$ are the straight lines separating
the sectors in Fig. 2. When crossing a Stokes line, the asymptotics of the solution to the differential
equation changes its nature.

\item Another approach is to solve the system (\ref{Schrgauge}). Were the constraint $\hat G \Psi = 0$ 
not imposed, the spectrum
would be $E_{mn} = n - m$ with the eigenfunctions $|nm\rangle = |n\rangle_x |m\rangle_y $. It is infinitely 
degenerate at each level.
 The constraint $G=0$ picks up only one representative of the set of eigenstates of $H$ with a given energy. 
For example, 
the zero energy state annihilated by $G$ is 
 \be
\label{Psi0}
\Psi_0 \ =\ \sum_{k=0}^\infty |2k,2k \rangle (-1)^k \frac {(2k-1)!!}{(2k)!!}\ .
 \ee
At large $k$, the coefficient is proportional to $1/\sqrt{k}$, i.e. the normalization integral for 
(\ref{Psi0}) diverges
logarithmically. 

Similarly, only one eigenstate is left at each energy level. The full spectrum is  discrete, 
\be
\label{Enravnon}
E_n = n
 \ee
 with positive, negative, or zero  integer $n$. 
It is unbounded both from
below and above. This notwithstanding, the spectral problem is well defined and the evolution operator 
 \be
\label{Evol}
{\cal K}(x', x)  \ =\ \sum_n \Psi_n^*(x')  \Psi_n(x)  e^{int}
 \ee
is unitary \footnote{See Ref.\cite{Robert} for detailed discussion of this and related issues.}. 

\end{itemize}

Comparing the results we obtained under two quantization procedures, one can make two observations. 
First, the spectrum is shifted 
by $1/2$. The ambiguity whether $E_n = n$ or $E_n = n + 1/2$ has the same nature as the well-known 
ordering ambiguity 
--- there are many different quantum problems having the same classical limit. The second observation
 is that, on 
top of the ordering ambiguity, there is in this case also another ambiguity associated with gauge choice.
 With  any gauge choice,
half of the spectrum involving either the states with negative or with positive  energies is lost.

 A lesson that can be drawn from this simple toy model is that, for gauge systems,  fixing the gauge 
classically and quantizing 
afterward may  be dangerous. Certain essential features of the spectral problem (\ref{Schrgauge}) may be lost.

\section{The potential $-(ix)^{2n+1}$. Classical dynamics.}

Having being equipped with necessary tools, we may proceed now with the analysis of the Hamiltonians of interest
written in Eqs.(\ref{2n+1},\ref{bezosc}). We will concentrate mainly on the Hamiltonian (\ref{bezosc}) without
the oscillator term in the potential.

Let first $n= 1$. 
Consider the complex Hamiltonian
  \be
 \label{Hz3}
{\cal H} \ =\ \frac {\pi^2}2 + i z^3\ 
  \ee
with $z = x + iy, \pi = p -iq$. Its real and imaginary parts are
  \be
\label{HG3}
 H &=& \frac {p^2 - q^2}2 + y^3 - 3yx^2 \ , \nn \\
 G &=&  -pq + x^3 - 3xy^2\ .
 \ee 

\vspace{-2cm}
\begin{figure}[h]
   \begin{center}
 \includegraphics[width=8.0in]{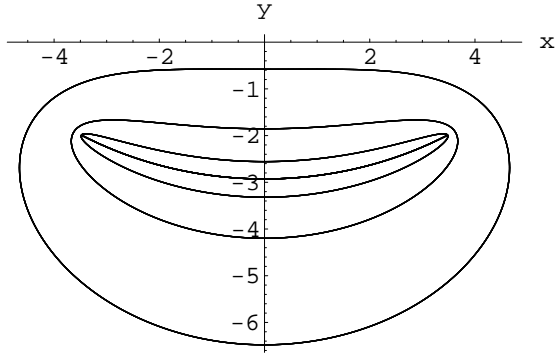}
        \vspace{-20cm}
    \end{center}
\caption{Family of trajectories with $E = 64$.}
\label{polorbity}
\end{figure}

Consider the dynamics of the system described by the Hamiltonian $H$ and the constraint $G$. 
It can be treated as a gauge system. The equations of motion follow from the Hamiltonian 
$H + \lambda G$, where $\lambda$ is the Lagrange multiplier. They have the form
 \be
\label{eqmotHG3}
\dot p = 6xy + 3\lambda(y^2 - x^2),\ \  \ \dot x = p - \lambda q, \ \ \
\dot q = 3(x^2 - y^2) + 6\lambda x y, \ \ \ \dot y = -q - \lambda p, \nn \\
 G = -pq + x^3 - 3xy^2 = 0 \ .
 \ee

The Lagrangian (\ref{Lcan}) is invariant up to a total derivative with respect to gauge 
transformations (\ref{gauge}) with time-dependent parameter $\alpha(t)$. 
To find the classical solutions, we need first to fix the gauge. A convenient {\it partial} gauge fixing
corresponds to the condition $\lambda(t) = 0$, in which case the equations are reduced to
 \be
\label{eqmotH3}
\dot p = 6xy,\ \ \dot x = p,\ \ \dot q = 3(x^2 - y^2), \ \ \dot y = -q , \ \ G=0\ .
 \ee

The solutions to (\ref{eqmotH3}) belong to two classes:
{\bf 1.} Runaway trajectories, which reach infinity at finite time. These are, for example, the trajectories
with initial conditions $x(0) = \dot x(0) = 0$. They run away in the positive $y$ directions. {\bf 2.} 
Besides, there are families
of closed orbits related to each other by gauge transformations (\ref{gauge}) with constant $\alpha$. For {\it positive} energies,
these families, depicted in Fig. 3, were found in Ref.\cite{BeBoMe}. This family has one distinguished member (one can call
it a {\it stem} trajectory): 
the trajectory which connects the turning points (the points where the monenta $p,q$ vanish) with the coordinates.
 \be
\label{turn}
y_* = - \frac {E^{1/3}}2,\ \ \ \ \ \ \ \ \ \ \ \ \ \ \ x_* = \pm \frac  {\sqrt{3} E^{1/3}}2 \ .
  \ee
(there is also the turning point $x = 0, y = E^{1/3}$, but the trajectories starting run away rather than coming back).
  
  Note that the families of trajectories with negative energies also exist (see Fig. 4). They stem from the trajectories
connecting the turning points    
 \be
\label{turn1}
y_{**} =  \frac {(-E)^{1/3}}2,\ \ \ \ \ \ \ \ \ \ \ \ \ \ \ x_{**} = \pm \frac  {\sqrt{3} (-E)^{1/3}}2 \ .
  \ee

\vspace{-3cm}
 \begin{figure}[h]
   \begin{center}
 \includegraphics[width=8.0in]{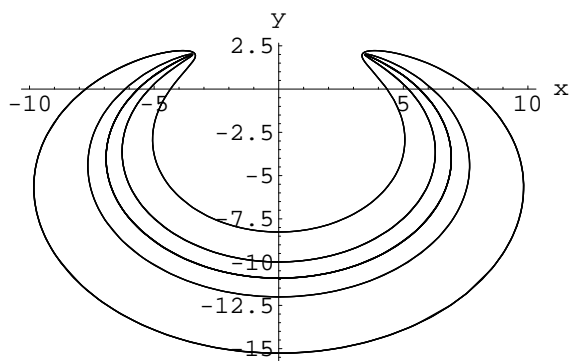}
        \vspace{-3cm}
    \end{center}
\caption{Family of trajectories with $E = -64$.}
\label{negorbity}
\end{figure}

Let us calculate for future purposes the action on these trajectories. Using the fact that the action for all
orbits belonging to one family is the same, one can write
 \be
\label{action}
 S = \oint (p dx + qdy) = 2 {\rm Re} \int_{z_1}^{z_2} \pi dz \ =\ 2 {\rm Re} \int_{z_1}^{z_2} \sqrt{2(E - iz^3)}  dz\ ,
 \ee
where $z_{1,2}$ are the turning points. For the trajectories of positive energies, the integral 
can be easily done by deforming the
contour such that it passes the origin 
\footnote{This result (in somewhat different normalization) was obtained in \cite{BeBoMe}.},
 \be
\label{S+} 
S_+ \ =\ 4 {\rm Re} \int_0^{z_*} \sqrt{2(E - iz^3)} dz = 4 {\rm Re} (z_*) \sqrt{2E} \int_0^1 \sqrt{1 - s^3}\, ds = 
\sqrt{6\pi} E^{5/6} \frac {\Gamma(4/3)}{\Gamma(11/6)}\ .
 \ee
 To calculate the action for negative energy orbits, one has to take into account the fact 
that the turning 
points are at the same time the branching
points of the integrand in (\ref{action}). For positive energies, the corresponding cuts  do not hinder the deformation of the
contour, but, for negative energies, they do. The cuts should be drawn such that the original path does not cross them. 
The deformed contour
also should avoid crossing the cuts. 
The corresponding structure of the cuts, the original and deformed contour are shown in Fig.5.
   \begin{figure}[h]
   \begin{center}
 \includegraphics[width=2.5in]{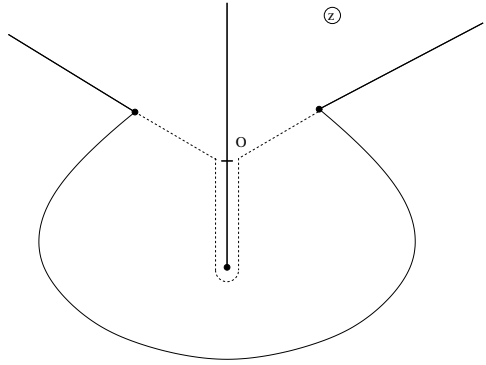}
    \end{center}
\caption{Analytic structure of $\pi_{\rm cl}(z)$ for negative energies. Solid line - original contour,
 dashed line - deformed contour, 
thick solid lines - the cuts.}
\label{razrezy}
\end{figure}

 It is clear from the figure that the deformed contour involves four pieces: {\it (i)} from the left turning point 
to the origin, {\it (ii-iii)} from the origin down the cut and up again, {\it (iv)} from the origin to the right turning
point. A simple analysis shows that the contribution of the parts (i-iv) involves an extra factor $\sin(\pi/6) = 1/2$ compared
to  the contribution of the parts (ii-iii).
 All together, the integral for $S_-$ involves an extra factor $ [1 + \sin(\pi/6)]/\cos(\pi/6) = \sqrt{3}$ 
compared to the integral (\ref{S+}) for $S_+$ 
with the same absolute value of energy. In other words,
  \be
\label{S-}
 S_- = 3 \sqrt{2\pi}   (-E)^{5/6} \frac {\Gamma(4/3)}{\Gamma(11/6)}
  \ee

\begin{figure}[h]
   \begin{center}
 \includegraphics[width=2.5in]{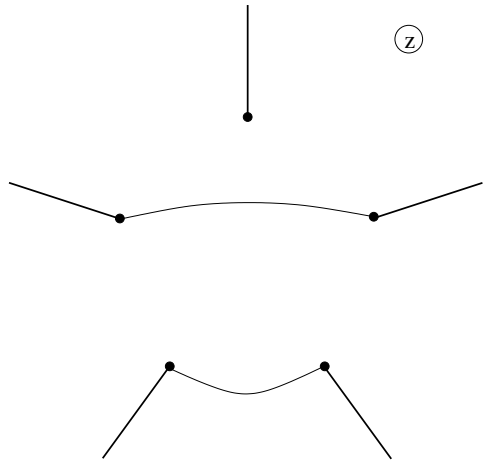}
    \end{center}
\caption{Turning points, cuts and stem trajectories for the potential $-iz^5$. Positive energies.}
\label{E+5}
\end{figure}

Consider now the complex Hamiltonian
\footnote{The sign of the potential corresponds to the convention (\ref{bezosc}) 
and  to the conventions of 
\cite{BenBoe,BeBoMe}. These conventions are convenient to make the physics of the systems (\ref{bezosc}) 
with different $n$ more similar.}
 \be
\label{Hz5}
{\cal H} \ =\ \frac {\pi^2}2 - iz^5
 \ee
 with real and imaginary parts
  \be
\label{HG5}
H = \frac {p^2 - q^2}2  + y^5 - 10 y^3 x^2 + 5 y x^4 \ , \nn \\
G = -pq - x^5 + 10y^2x^3 - 5xy^4 \to 0\ .
 \ee

\begin{figure}[h]
   \begin{center}
 \includegraphics[width=3.0in]{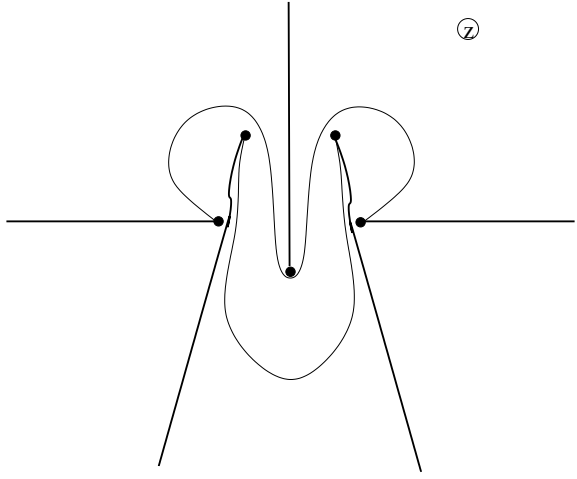}
    \end{center}
\caption{The same for negative  energies. The cuts are drawn not to interfere with the paths.}
\label{E-5}
\end{figure}

 Again, there are runaway trajectories taking a finite time to reach infinity in the positive $y$ direction.
Besides, there are {\it four} families of closed orbits: two families with positive energies and two 
families with negative
energies. The structure of the turning points, associated cuts and the stem trajectories 
connecting the turning points is shown schematically in Figs. 6,7 for 
positive and negative energies, respectively. 
 Let us find the classical action at these trajectories.
For positive energies,
 \be 
\label{S+updown}
S_+^{\rm up} &=& 4 \cos \frac \pi{10} \sqrt{2} \int_0^1 \sqrt{1-s^5} \, ds\,  E^{7/10}\ =\ 2 \sqrt{2\pi} 
\cos \frac {\pi}{10} \, \frac {\Gamma(6/5)}{\Gamma(17/10)} \, E^{7/10}  \ , \nn \\
S_+^{\rm down} &=& 4 \cos \frac {3\pi}{10} \sqrt{2} \int_0^1 \sqrt{1-s^5}\, ds\,  E^{7/10}\,  \ =\ 2 \sqrt{2\pi} 
\cos \frac {3\pi}{10}\,  \frac {\Gamma(6/5)}{\Gamma(17/10)}  E^{7/10} \ .
 \ee
   For negative energies,
 \be 
\label{S-updown}
S_-^{\rm up} &=&  2 \sqrt{2\pi} \left(1 + 2\sin \frac {3\pi}{10}  + \sin \frac \pi{10} \right) \, 
 \frac {\Gamma(6/5)}{\Gamma(17/10)}  (-E)^{7/10} \ , \nn \\
S_-^{\rm down} &=&  2 \sqrt{2\pi} 
\left(1 + \sin \frac {3\pi}{10} \right) \frac {\Gamma(6/5)}{\Gamma(17/10)}\,  (-E)^{7/10} \ .
 \ee
The superscript ``up'' in Eq. (\ref{S-updown}) refers to the upper trajectory in Fig. 7 going between the points
$e^{11i\pi/10}$ and $e^{-i\pi/10}$. The result for $S_-^{\rm down}$ is obtained in the same way as the result
 (\ref{S-}), with the factor $1 + \sin(\pi/6)$  being replaced by  $1 + \sin(3\pi/10)$. When deforming the 
contour for the upper trajectory, we find, in addition to the parts composing the deformed contour of the lower
trajectory and giving the factor $1 + \sin (3\pi/10)$, also two extra pieces  with the contribution $\sim$
$\sin (3\pi/10) + \sin (\pi/10)$. The origin of all these factors can be clearly seen, if  deforming
 the contour and the cuts in the way shown in Fig. 8. All the pieces (of nonzero length) connect the branching points
to the center of the pentagon $z=0$.  
  \begin{figure}[h]
   \begin{center}
 \includegraphics[width=4.0in]{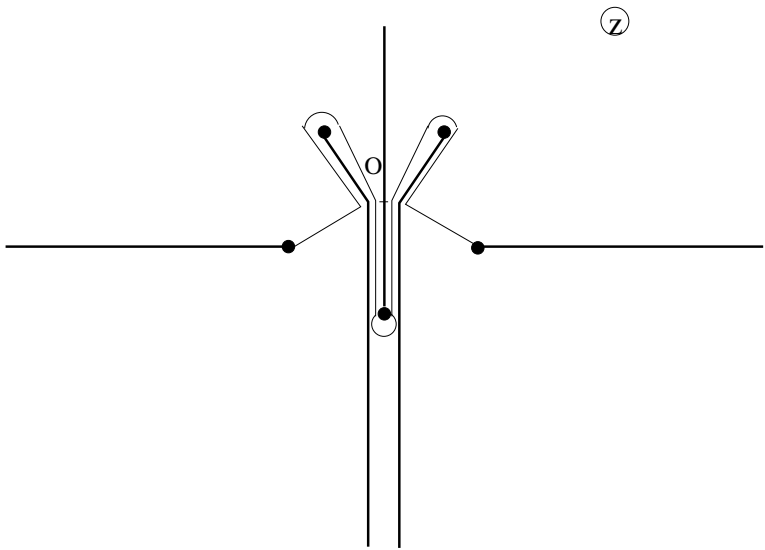}
    \end{center}
\caption{The deformed upper contour wiggling around the deformed cuts. }
\label{deform5}
\end{figure}

 By the same token, for the potential $-(iz)^{2n+1}$, there are $2n$ families of the trajectories: $n$ families
with  positive energies and $n$ families with  negative energies. 

As we have seen, the classical dynamics of the system with the potential $-(iz)^{2n+1}$ is similar 
in many respects to the complex oscillator dynamics: a distinct feature of both systems are the families of 
closed orbits with positive and negative energies, the members of one family being interrelated by gauge
transformations. There are also two important differences. First,  
the system $-(iz)^{2n+1}$ involves besides closed orbits also singular runaway trajectories. Second, 
for the complex oscillator, the stem trajectories for the families of orbits could be conveniently
obtained by fixing the gauge $y=0$ or $x=0$. But for the system $-(iz)^{2n+1}$, this is not true.
 To begin with, the stem trajectories displayed above are essentially complex. This observation is not yet
sufficient, however, because it does not exclude a conceivable in principle  possibility that the trajectories can be 
put onto the real (or imaginary) axis by a complicated gauge transformation (\ref{gauge}) with nontrivial  $\alpha(t)$.
 
 Let us find out what happens if we {\it do} fix the gauge $y=0$ for the 
system (\ref{HG3},\ref{eqmotHG3}).
From $G=0$, we deduce  $q = x^3/p$ and hence the Hamiltonian is reduced to 
 \be
\label{Hstar}
 H^* \ =\ \frac {p^2}2 - \frac  {x^6}{2p^2} \ .
 \ee
 The corresponding equations of motion
\be
\label{eqmotstar}
 \dot x = p + \frac { x^6}{p^3} \ , \ \ \ \ \ \ \ \ \ \ \ \ \ \dot p = \frac {3 x^5}{p^2}
 \ee
follow from (\ref{eqmotHG3}) with $\lambda = -x^3/p^2$.
 We see now that the reduced Hamiltonian (\ref{Hstar}) is neither positive nor negative definite and involves
{\it only} runaway trajectories. Closed orbits have disappeared ! This is another manifestation of the fact
discussed in the previous section that fixing the gauge at the classical level is not an innocent procedure
and may lead to a loss of important dynamic features. For the complex oscillator with the gauge choice $y=0$, 
 half of the orbits (the orbits with negative energies) were lost. For the system $-(iz)^{2n+1}$,  {\it all} closed orbits 
are lost and we are left only with runaway solutions.
  
 Let us discuss the relationship of the Hamiltonian (\ref{Hstar}) to another Hamiltonian obtained
from (\ref{Hz3}) by  a non-unitary rotation technique in the spirit of \cite{Most}.
Let us multiply the potential by a coupling constant $g$, $ix^3 \to igx^3$, and find an operator
$R$ such that the rotated Hamiltonian
$\tilde H = e^R ( p^2/2 + igx^3)  e^{-R}$ be manifestly real. Then $R$ can be presented as an infinite series over the
coupling constant,
 $$R = - \frac {gx^4}{4p} + O(g^3) $$ and \cite{Most2} (see also sect. V of Ref.\cite{BeBroJo})
 \be
\label{Hrotate}
  \tilde H \ =\ \frac {p^2}2 + \frac  {3g^2 x^6}{8p^2} + O(g^4) \ .
 \ee
 We see that the  $H^*$ and $\tilde H$  have  similar structure, but the coefficients differ.
 This does not represent
a paradox because $\tilde H$, in  contrast to $H^*$, involves the whole infinite series in $g$. Anyway, 
all the terms in this series are nonlocal, and one cannot obtain from this, say,  the spectrum
 of quantum Hamiltonian
as a perturbative series in $g$. The nonunitary rotation techniques is better suited to the problems 
like (\ref{2n+1}),
where all the terms in the perturbative series for $\tilde H$ are local.       
 
 Coming back to  fixing the gauge with the condition $y = 0$, it does not work well  also for the mixed 
system  (\ref{2n+1}),
however small $g$ is. The extra piece in $H^*$ is still nonlocal and singular at  
the turning point of the unperturbed oscillator trajectory where momentum $p$ vanishes. 
As a result, the trajectory 
does not {\it turn} there, but rather stumbles and runs away.

\section{Quantum dynamics.}

Let us discuss  now  quantum dynamics of the Hamiltonians (\ref{Hz3}), (\ref{Hz5}). Consider Eq.(\ref{Hz3}) first.
In Sect. 2, we outlined two regular ways to quantize gauge systems: {\it (i)} resolving the constraint(s) at the classical
level and quantizing afterward, and  {\it (ii)} solving the system of differential equations (\ref{Schrgauge})
 with proper boundary conditions. 

To resolve the constraints classically, one has to fix the gauge. Unfortunately, as we have just seen, it is difficult 
to find a clever way to do it in our case. A natural gauge fixing leads to the problem involving {\it only} runaway trajectories.
 This means trouble and, indeed, for the highly nonlocal and  not positive definite Hamiltonian (\ref{Hstar}),
 one cannot formulate
a well-defined quantum problem with a unitary evolution operator. 

Another approach is to solve the system (\ref{Schrgauge}). This is a nontrivial numerical problem. 
Indeed, one-dimensional spectral problems can be easily solved with Mathematica, but in this case the problem is
essentially two-dimensional, which is much trickier. What is even more important, the operators $H$ and $G$ in (\ref{Schrgauge})
 are not elliptic, as usual, but hyperbolic. It is not thus evident that a reasonable
solution to this problem exists... We will discuss this question somewhat more in the last section, but, basically, 
we leave
it for future studies.

There is, however, a way to define a consistent spectral problem related to the Hamiltonian (\ref{Hz3}) \cite{BenBoe}.
Forget for a moment all what was said above about complexification and consider the Schr\"odinger equation at the real
axis,
 \be
\label{Schrx}
\left[- \frac 12 \frac \partial {\partial x^2} + ix^3 \right] \Psi \ =\ E \Psi \ .
\ee
 with the condition that the wave function falls down at $x = \pm \infty$. It is convenient  to pose the problem
not on the whole line $(-\infty, \infty)$, but on the half-line $(0, \infty)$. One can do it by exploiting 
 the $PT$-symmetry of the potential
(the property $V(-x) = V^*(x)$). It  dictates that for any solution $\Psi(x)$ of Eq.(\ref{Schrx}),
the function $\Psi^*(-x)$ is also the solution with the same eigenvalue. The functions
 \be
\label{psipm}
 \Psi_\pm (x) \ =\  \Psi(x) \pm \Psi^*(-x)
 \ee
with the symmetry properties  $\Psi_+(-x) = \Psi_+^*(x)$ and  $\Psi_-(-x) = -\Psi^*_-(x)$ 
also satisfy this equation. We are hence allowed to consider the equations for $PT$-even function $\Psi_+(x)$ and $PT$-odd
function $\Psi_-(x)$ separately. In this case (in contrast, e.g. to the standard oscillator problem), the equation
for $\Psi_-(x)$ does not give anything new. Indeed, one can make a $PT$-odd function out of a $PT$-even one by simply 
multiplying the latter by $i$. A generic solution to (\ref{Schrx}) is obtained by multiplying a $PT$-even solution by an 
arbitrary
complex factor.

The condition  $\Psi(-x) = \Psi^*(x)$ means that $\Psi(0)$ is real  while $\Psi'(0)$ is imaginary.
By turning  computer on, everybody can be convinced that the equation (\ref{Schrx}) with the boundary conditions
  \be
\label{boundcond}  
{\rm Re} \left( \frac {\Psi'(0)}{\Psi(0)} \right) = 0,\ \ \ \Psi(\infty) = 0 
   \ee
has solutions at real positive discrete values of $E$. The remarkable fact is that these values are very 
close to {\it semiclassical} energies associated with the family of the closed orbits in Fig. 3 obtained from 
the quantization
condition
 \be
 \label{semicl}
S(E_k) \ =\ \pi(2k+1)\ ,
  \ee
with the function $S(E)$ being given by Eq.(\ref{S+}). When $k \to \infty$, the spectral values
extracted from Eqs.(\ref{Schrx},\ref{boundcond}) and the semiclassical values extracted from Eqs.(\ref{S+},\ref{semicl})
rapidly converge.  The exact and semiclassical values for $E_k$ for  first few levels \cite{BenBoe}  are shown in Table 1. 
\vspace{.3cm}
$$
\begin{array}{c|c|c|c|c}
k & 0 & 1 & 2 & 3 \\
\hline
E^k_{\rm exact} & 0.763 & 2.711 & 4.989 & 7.465 \\
\hline
E^k_{\rm semicl} & 0.722 & 2.698 & 4.980 & 7.458 
 \end{array}
$$
 
\vspace{.2cm}

 \centerline{{\sl Table 1.} Exact and semiclassical spectra for the potential $ix^3$.}

\vspace{.2cm}

Once the solution is obtained, one need not to stay on the real axis. Actually, the solution can be continued
analytically to complex values of the argument $z$ in the regions
 \be
 \label{Argreg}
\left|   {\rm Arg}(z)  + \frac \pi {10} \right| \leq \frac \pi 5\ , \ \ \ \ \ \ \ 
 \left|   {\rm Arg}(z) -  \frac {11 \pi} {10} \right| \leq \frac \pi 5 \ .
 \ee
In other words, the spectral problem
   \be
\label{Schrs}
\left[- \frac 1{2\Phi^2} \frac \partial {\partial s^2} + is^3 \Phi^3\right] \Psi \ =\ E \Psi\ ,  \nn \\
\left. {\rm Re} \left( \frac {\partial \Psi /\partial s}{\Phi \Psi} \right) \right|_{s=0} \  = \ 0,\ \ \ \ 
\Psi(\infty) = 0 \ ,
\ee
with $\Phi = e^{i\alpha}$, still has a solution when $\alpha$ lies within the interval (\ref{Argreg}), and the spectral
values are exactly the same as for the problem (\ref{Schrx},\ref{boundcond}). When 
 \be
\label{intvnizu}
-7\pi/10 < \alpha < -3\pi/10 \ ,
 \ee
 the spectrum is continuous: any  positive or negative energy is acceptable. This is especially 
clearly seen for $\alpha = -\pi/2$
(meaning $\Phi = -i$). The problem (\ref{Schrs}) is then reduced to
   \be
\label{Schrs1}
\left[- \frac 1{2} \frac \partial {\partial s^2} + s^3 \right] \Psi \ =\ -E \Psi  \nn \\
 \Psi(0) = 1, \ \ \ \ \ \ \ \Psi(\infty) = 0, \ \ \ \ \ \ \ \ \ {\rm Im} \, \left[ \Psi'(0) \right]  = 0\ .
\ee
The real part of $\Psi'(0)$ is not fixed, however, and tuning this parameter, one can obtain the solution 
dying at infinity at any energy. 
\footnote{By modifying the spectral problem by, for example, imposing the conditions  $\Psi(0) = \Psi(\infty) = 0$ \ instead of
(\ref{boundcond}), one can force the spectrum to be discrete and negative definite. But the condition $\Psi(0) = 0$ 
is artificial and has no physical motivation. In particular, the discrete negative definite 
spectrum thus obtained has nothing
to do with the semiclassical spectrum (\ref{E-semicl}). } 
A numerical analysis shows that it is true in the whole interval (\ref{intvnizu}).

On the other hand, for $\pi/10 < \alpha <9\pi/10$  the problem (\ref{Schrs}) has no solution whatsoever: the 
spectrum is empty. 
\footnote{ If lifting the requirement that the wave function dies away at infinity, 
the spectrum would again become continuous.} This is all illustrated in Fig. 9. 

 \begin{figure}[h]
   \begin{center}
 \includegraphics[width=3.5in]{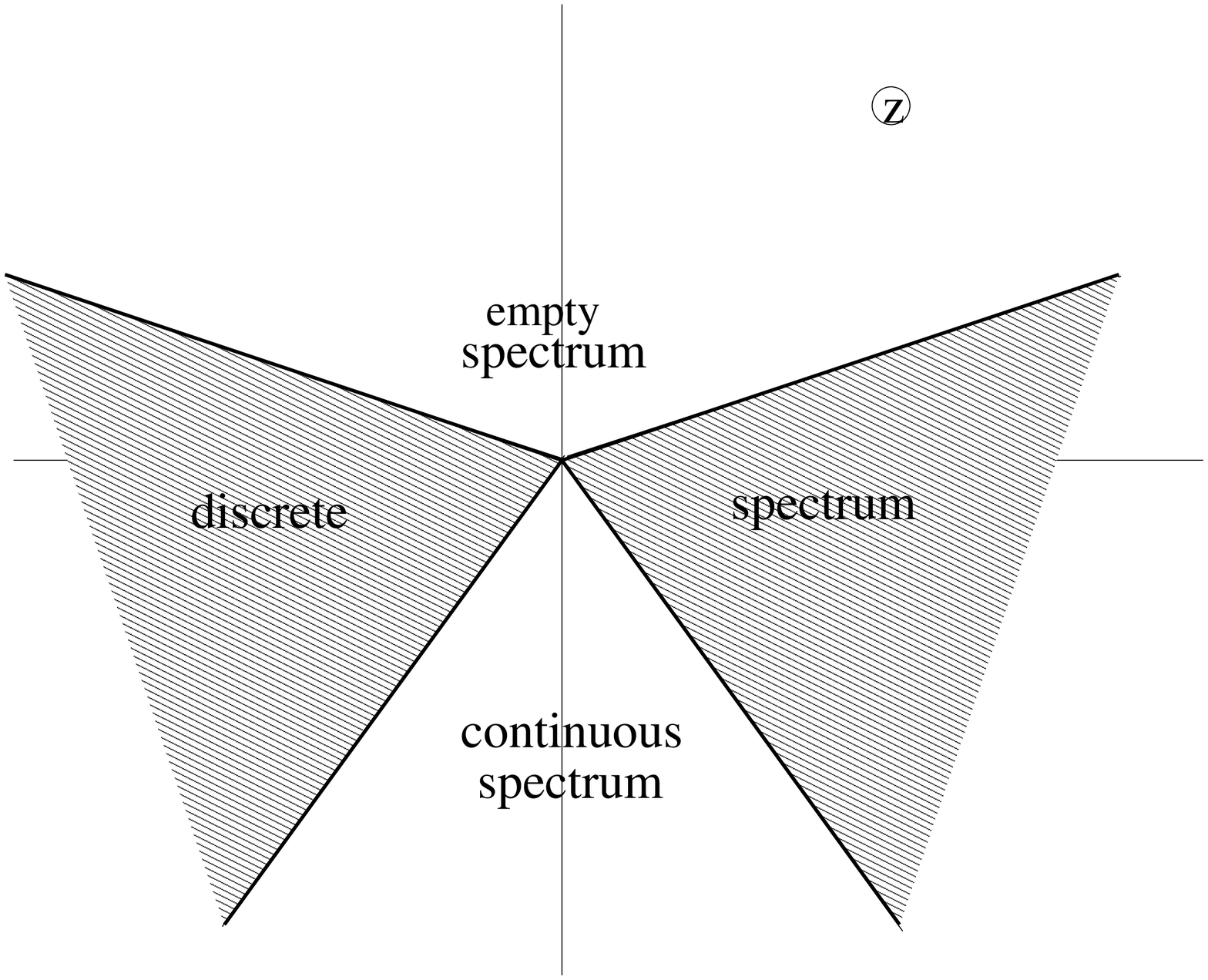}
    \end{center}
\caption{Spectral problem (\ref{Schrs}) in the complex $z$ plane.}
\label{sektora3}
\end{figure}

The system of the lines separating the sectors in Fig.9 form together with the positive imaginary axis the system
of the asymptotes of the Stokes lines of the Schr\"odinger equation with the potential $iz^3$. Generically, for a polynomial
potential of order $n$, such system involves $n+2$ lines forming equal angles $2\pi/(n+2)$ \cite{Stokes}. 

The spectral problem (\ref{Schrs}) corresponds to the family of the classical orbits in Fig.3 with {\it positive} energies.
As we have seen (in Fig. 4), there are also orbits with negative energies. Using the result (\ref{S-}), 
it is not difficult
to find the corresponding semiclassical energies,
 \be
 \label{E-semicl}
E_k \ =\ - \left[ \frac {(2k+1) \Gamma(11/6)}{\Gamma(4/3)} \right]^{6/5} \left( \frac \pi{18} \right)^{3/5}\ .
  \ee 
One may suggest that a spectral problem should exist for which Eq.(\ref{E-semicl}) would represent a semiclassical
approximation. However, no such problem is known.
\footnote{And here is an important difference with the complex oscillator problem discussed in Sect. 2, where the spectral
problem with the spectrum $E_k = -k - 1/2$ was perfectly well defined.} 
 At least, it is not known in the standard form of boundary
problem for some differential operator. One still can calculate the ``exact spectrum'' of such nonexisting
 (or very well hidden) problem by calculating corrections to the result (\ref{E-semicl}) and representing
$E^k_{\rm exact}$ as a series in semiclassical parameter  $\sim 1/S_{\rm cl} $. 
As this series is probably 
asymptotic, this method gives an intrinsic uncertainty in the spectrum $\sim \exp\{-C S_{\rm cl}\}$. However,
the closeness of exact energies of positive energy states and their semiclassical approximations (see Table 1) 
and  the calculations of higher order corrections in 
 \cite{BeBoMe}    suggests that  this uncertainty is not large even for the ``sky state''  
in Eq. (\ref{E-semicl}) with
$k = 0$ and $S_{cl} = \pi$. It rapidly decreases with increase of  $k$.

Consider now the Hamiltonian (\ref{Hz5}). Again, one can solve the Schr\"odinger equation with the potential 
$-ix^5$  at  the real axis  with boundary  conditions $\Psi(\pm \infty) = 0$ and find a discrete
 spectrum with real positive energies. As is seen from Table 2, these exact
 energies are very close to semiclassical values determined from the quantization condition
 \be
\label{kvantov}
S_+^{\rm up} \ = \ \pi(2k+1) \ ,   
  \ee
where $S_+^{\rm up}$ given in Eq.(\ref{S+updown}) is evaluated for the {\it upper} trajectory in Fig. 6.

\vspace{.3cm}
$$
\begin{array}{c|c|c|c|c}
k & 0 & 1 & 2 & 3 \\
\hline
E^k_{\rm exact} & 0.710 & 2.660 & 5.458 & 8.788 \\
\hline
E^k_{\rm semicl} & 0.543 & 2.608 & 5.410 & 8.750 
 \end{array}
$$
 
\vspace{.2cm}

 \centerline{{\sl Table 2}. Exact and semiclassical spectra for the potential $-ix^5$.}

\vspace{.2cm}

As we see, semiclassical approximation works somewhat worse in this case than for the potential $ix^3$.
But it works. 

We can now leave the real axis and solve the spectral problem
   \be
\label{Schrs5}
\left[- \frac 1{2\Phi^2} \frac \partial {\partial s^2} - is^5 \Phi^5 \right] \Psi \ =\ E \Psi \ , \nn \\
\left. {\rm Re} \left( \frac {\partial \Psi /\partial s}{\Phi \Psi} \right) \right|_{s=0} \  = \ 0,\ \ \ \ 
\Psi(\infty) = 0 \ ,
\ee
with $\Phi = e^{i\alpha}$.  The solution {\it with the same spectrum} exists for 
 \be
\label{interv5u}
\left|\alpha - \frac \pi{14} \right| \leq \frac \pi 7, \ \ \ \ \ \ {\rm or} \ \ \ \ \ \   
\left|\alpha  - \frac {13\pi} {14} \right| \leq \frac \pi 7 \ .
  \ee
For 
 \be
\label{interv5d}
\left|\alpha + \frac {3\pi}{14} \right| \leq \frac \pi 7, \ \ \ \ \ \ {\rm or} \ \ \ \ \ \  
\left|\alpha  - \frac {17\pi} {14} \right| \leq \frac \pi 7 \ ,
  \ee
the solution still exists, but the spectrum is different. Its semiclassical approximation comes not from
the quantization condition (\ref{kvantov}), but rather from the quantization condition
  \be
\label{kvantov1}
S_+^{\rm down} \ = \ \pi(2k+1)    
  \ee
derived for the {\it lower} stem trajectory in Fig.6. 
The exact and semiclassical energy values for this case are given in Table 3.

$$
\begin{array}{c|c|c|c|c}
k & 0 & 1 & 2 & 3 \\
\hline
E^k_{\rm exact} & 1.163 & 5.234 & 10.795 & 17.428 \\
\hline
E^k_{\rm semicl} & 1.080 & 5.186 & 10.759 & 17.400 
 \end{array}
$$
 
\vspace{.2cm}

 \centerline{{\sl Table 3}. Exact and semiclassical spectra for the potential $-iz^5$ in the region 
(\ref{interv5d}).}

\vspace{.2cm}

Finally,   for  $-9\pi/(14) < \alpha < -5\pi/(14) $, the spectrum is continuous while, for  
$3\pi/(14) < \alpha < 11\pi/(14) $, the spectrum 
is empty. The corresponding regions in the complex $z$ plane 
are displayed 
in Fig.10. 
  \begin{figure}[h]
   \begin{center}
 \includegraphics[width=3.5in]{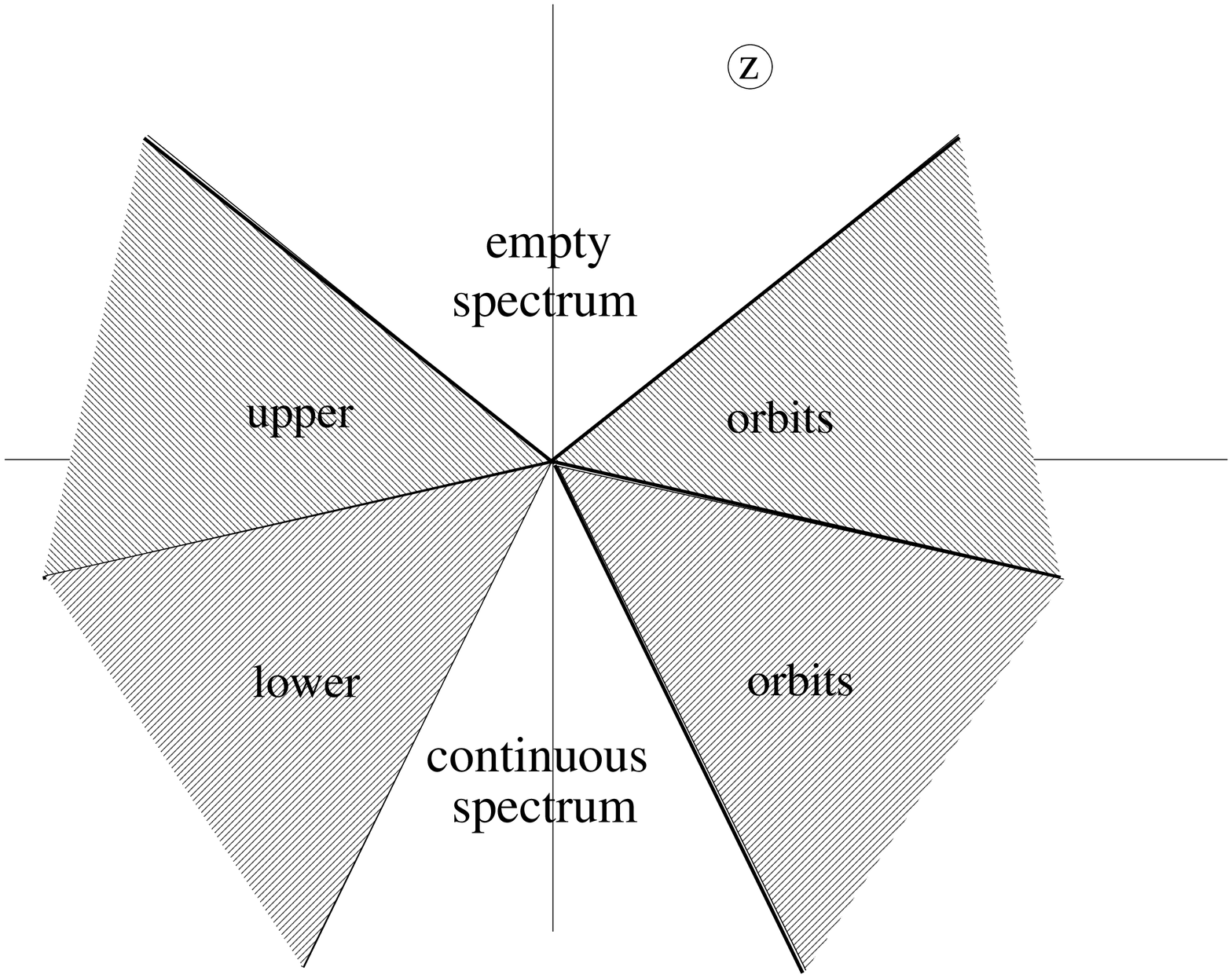}
    \end{center}
\caption{Spectral problem (\ref{Schrs5}) in the complex $z$ plane.}
\label{sektora5}
\end{figure}

For the Hamiltonian (\ref{2n+1}), there are $n$ different nontrivial spectral problems with discrete spectrum 
defined in the sectors
 \be
 \label{sectn}
\left| \alpha + \frac {(2n-1) \pi }{2(2n+3)}  - \frac {2\pi m}{2n + 3} \right| \leq \frac \pi {2n+3}
\ \ \ 
 {\rm or\  \ \ \ \ \ \ \ \ \ mirror\ image}\ ,
 \ee
$m = 0,\ldots, n-1$. 
\footnote{There are also asymmetric spectral problems. One can, for example, go from infinity to zero along the line
$\alpha = 17\pi/14$ and to infinity from zero along the line $\alpha = \pi/14$. But such problems have 
complex eigenvalues \cite{Shin} and we are not considering them.}
They correspond to $n$ different families of classical orbits with {\it positive}
 energies for the potential $-(iz)^{2n+1}$. The problem studied in details in Refs.\cite{BenBoe,BeBoMe}
was defined in the sector $m = 0$.
\footnote{
 In Ref. \cite{BeBoMe}, the problem with $m=1$ was also considered. It was represented as the problem 
with the potential $V(x) = x^4(ix)^\epsilon$. But the results for the spectrum were given there only 
for negative $\epsilon$.}    
We concentrate in this paper on imaginary potentials like in Eq.(\ref{bezosc}). But one can equally well 
\cite{BenBoe,BeBoMe} consider the potentials 
 \be
\label{2n}
 V(z) = -(iz)^{2n}
 \ee
 For the quartic potential $\sim - z^4$, 
there are four turning points, two sets of symmetric classical positive energy orbits and a 
corresponding spectral problem defined
in the sector $|\alpha - \pi/6| \leq \pi/6$ and its mirror images. For the  potential $\sim z^6$ we have, besides
the standard spectral problem on the real axis, also a nontrivial problem in the sector $|\alpha - \pi/4| \leq
\pi/8$, etc. For generic $n$, the potential (\ref{2n}) admits $n/2$ different spectral problems when $n$ is even
and $(n+1)/2$ different spectral problems when $n$ is odd.

 The presence of several different quantum problems associated with a given classical potential seems to be natural
in view of our analysis for the complex oscillator, where two different spectral problems exist. However, 
it might appear surprising in the framework of Mostafazadeh's approach where the crypto-Hermitian Hamiltonian is obtained
by a nonunitary rotation out of Hermitian $\tilde H$  representing a quite definite series in $g$ 
and hence the spectrum of $H$ and
of $\tilde H$  represents a quite definite series in $g$. For example, the ground state energy
of the system 
 \be
 \label{igx5}
 H \ =\ \frac {p^2 + x^2} 2 - ig x^5
 \ee
is 
\be
\label{Evacx5}
  E_0 \ =\ \frac 12 + \frac {449g^2}{32} + O(g^4)\ .
 \ee
The resolution of this  paradox is the following. Seemingly, 
only {\it one} of the spectral problems (\ref{Schrs5}) associated
with the Hamiltonian (\ref{igx5}), the problem defined in the sector including the real axis, can be safely 
treated in the framework of
Mostafazadeh's approach. The ground state energy   is  plotted in Fig. 11 as a function of  $g$.
Indeed, the spectrum tends to the oscillator spectrum when $g \to 0$. It is not seen on the plot, but  for very 
small $g$ , starting from $g \approx .01-.02$, the numerical values of the energies agree with  the perturbative evaluation
(\ref{Evacx5}).  

 \begin{figure}[h]
   \begin{center}
 \includegraphics[width=3.5in]{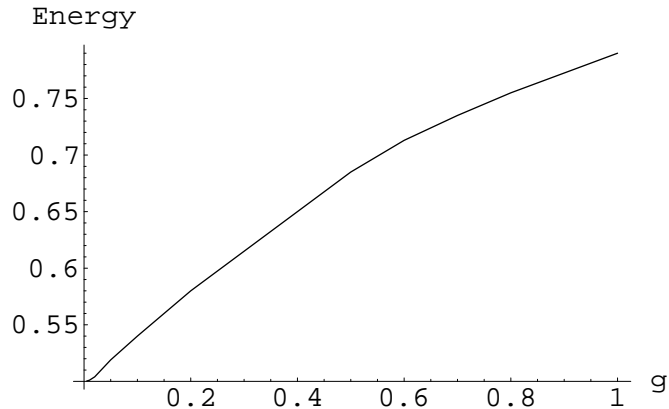}
    \end{center}
\caption{Ground state energy of (\ref{igx5}) as a function of $g$. Upper region.}
\label{up0}
\end{figure}

The solution for another spectral problem at the vicinity of the rays $\alpha = -3\pi/14$ and 
$\alpha =  17\pi/14$  behaves in a different
 and rather unexpectedly different way. For very small $g$, the spectrum is transformed, indeed,
 to the oscillator spectrum, but this transformation
occurs in a very  nontrivial manner. When $g$ goes down, the energies of all the states go down in such a way that
the energy of the ground state gets closer and closer to the energy of the first excited state. 
At some critical value of the coupling $g_* \approx .03717$, their energies coincide,
$$ E_0(g_*) \ = \  E_1(g_*) \ \approx  \ .484 \ . $$
  At still lower values of $g$, the energies should become complex. 
 On the other hand, the second excited state goes down and down with 
decreasing of $g$ and approaches the ground state oscillator energy without adventures, $E_2(g \to 0) \to  1/2$ (see Fig. 12).

 \begin{figure}[h]
   \begin{center}
 \includegraphics[width=3.5in]{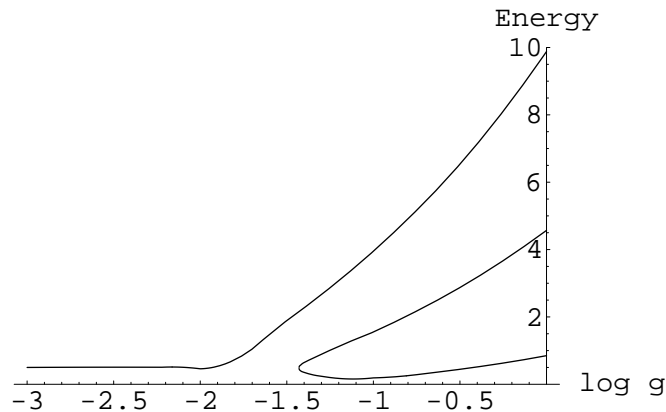}
    \end{center}
\caption{Three first levels of (\ref{igx5}) as a function of $g$. Lower region.}
\label{down012}
\end{figure}

 To be more precise, there are no adventures in a sense that there is no phase transition and the state exists
 at any $g$ and has real energy. But the asymptotics is reached only at rather small couplings.
The energy of the second excited state  finds itself at the vicinity of $E = 1/2$ only at $g \approx .01$. Now, $E_2(.01) 
\approx
.46$ and does not coincide with the perturbative expansion (\ref{Evacx5}). It is not excluded that at still smaller 
values of coupling, $g \approx .001$, the perturbative asymptotics (\ref{Evacx5}) finally shows up. To see whether 
it is true or
not,  a more careful numerical study is required.

The third and the fourth excitations of the Hamiltonian (\ref{igx5}) coalesce 
and their energies cease to be real  at 
$g_{**} \approx .007$ (the energy is $E_{**} \approx 1.37$ at this point), 
while the fifth excitation approaches the first oscillator excitation $E = 3/2$ 
at very small values of $g$.
One can suggest that this pattern holds also further up: the 6-th and the 7-th excitations  of the mixed Hamiltonian
coalesce and their energy becomes complex at some very small $g_{***}$, while the 8-th  excitation approaches the second 
oscillator
excitation $E = 5/2$, and so on. We thus observe an infinite sequence
 of ``phase transitions'' in the coupling.
\footnote{ This kind of  transition when a pair of real eigenstates of
 a boundary problem coalesce and become complex
 is a known phenomenon \cite{Heiss}. Its essence is clearly seen in 
 a trivial example. The matrix
$$ A =  \left( \begin{array}{cc} 1 & 1 \\ \alpha & 1 \end{array} \right) $$
has a pair of close real eigenvalues at small positive $\alpha$ 
and a pair of complex conjugated eigenvalues 
for $\alpha <0$. When $\alpha = 0$ (the exceptional point),
 the matrix represents a nondiagonalizable Jordan block.  

  An infinite set of such  transitions in the
parameter $\epsilon$ for the problem $V(x) = x^2(ix)^\epsilon$ was observed in \cite{BenBoe}.  
We observed a similar phenomenon in a completely different physical context:  
it happens that some {\it domain wall} solutions in supersymmetric gauge theories disappear 
when mass of the matter fields exceeds certain critical values \cite{Vesel}.}

This analysis shows that the Hamiltonian (\ref{igx5}) is crypto--Hermitian  for all couplings in the upper sectors in Fig. 10, but,
in the lower sectors, it is true only for not too small $g > g_\star$. When $g < g_\star$, a pair of complex 
conjugate eigenvalues should appear. For $g < g_{\star\star}$, there are two such pairs, etc. 
It would be very interesting to see these complex eigenvalues explicitly. Unfortunately, it is not so easy to do it with our methods 
- the spectral problems 
of the type (\ref{Schrs}, \ref{Schrs5}) make sense only for real energies --- the boundary condition
$$
\left. {\rm Re} \left( \frac {\partial \Psi /\partial s}{\Phi \Psi} \right) \right|_{s=0} \  = \ 0
$$ 
was derived under the assumption that $\Psi(z)$ and $\Psi^\star(-z)$ satisfy the same Schr\"odinger equation, which is only true
when $E$ is real. A special study of this issue is required.

\section{Discussion and Outlook.}

Crypto-Hermitian systems have many common features with the systems involving higher derivatives. In both cases,
Hermiticity of the Hamiltonian and unitarity of the evolution operator seem to be lost, but, if treating the problem properly, 
it is often restored. There exist also a more concrete relationship between two kind of systems.
 We have seen that the real part $H$ of the complexified Hamiltonian [see Eqs.(\ref{Hosc}),
(\ref{HG3})]  is never positive definite and may give rise to ghosts. The same is true for higher-derivative theories. Actually,
the canonical Hamiltonians of the latter have a rather similar form with not positive definite kinetic term \cite{Robert}.
The resemblance between the supersymmetric system analyzed in Ref.\cite{Robert} and the problem considered here is even more 
striking.
 A system of the type (\ref{HG3}) involves besides $H$ the integral of motion $G$, and we are interested with the sector $G=0$.
 The system studied in Ref.\cite{Robert} (the bosonic part of its Hamiltonian is 
 \be
\label{HbosRob}
 H = pP - DV'(x)\ ,
 \ee
 where $(p,x)$ and $(P,D)$ are two pairs
of canonic variables and superpotential $V(x)$ is an arbitrary function) also possesses an extra integral of
motion $N = P^2/2 - V(x)$. In the sector with a particular value of $N$ (including $N=0$) , the spectrum is discrete involving
positive and negative energies.

 The latter is also true for the spectrum (\ref{Enravnon}) of complexified oscillator when the constraint $G = 0$ is 
imposed on the quantum states  as in Eq.(\ref{Schrgauge}). The Dirac quantum problem (\ref{Schrgauge}) 
is more naturally posed than other
quantum problems associated with the classical system in hand. This problem is easily solved
in the oscillator case, but, for the potential $iz^3$, this is a {\it difficult} numerical
problem, and we leave it for further studies. One can speculate that its spectrum  involves
 positive and negative energies, as the spectrum
of the complexified oscillator and the spectrum of the Hamiltonian (\ref{HbosRob}) do. However, it is an open question at present 
whether the problem (\ref{Schrgauge}) {\it makes sense} for potentials more complicated than $z^2$. 
As we have seen in \cite{Robert},
the Hermiticity of the Hamiltonian (\ref{HbosRob}) and the unitarity of the corresponding evolution operator are corollaries 
of the fact that  classical trajectories of this system are benign enough: there are no collapsing or runaway trajectories where
a singularity is reached at finite time.
 On the other hand, for the systems (\ref{2n+1}), runaway classical trajectories exist. For sure, not all the trajectories
 associated with the systems (\ref{2n+1}) are
runaway trajectories. There are also closed orbits, and a hope that the 
problem (\ref{Schrgauge}) is well defined is associated
with their existence. The presence of runaway trajectories may spoil the brew, however. 
 
Runaway trajectories {\it definitely} spoil the brew for the quantum problems obtained by resolving 
the gauge constraint $G=0$ at the classical level.
This procedure gives benign sensible Hamiltonians for the complexified oscillator.
However, the Hamiltonian (\ref{Hstar}) thus obtained is not Hermitian and unitarity is lost too. 

There are, however, Hermitian and unitary quantum problems associated with the Hamiltonians (\ref{2n+1}) and 
(\ref{bezosc}). For one of such problems corresponding to
the potential $x^2/2 - igx^5$ in the sectors below the real axis, 
we discovered a rather interesting and nontrivial phenomenon: when the coupling constant $g$ 
is decreased, certain quantum states coalesce and disappear from the physical (real energy) spectrum. The number of such phase transitions 
is infinite, which reminds an infinite number of phase transitions in $\epsilon$ for the potential $x^2(ix)^\epsilon$
observed in \cite{BenBoe}. 
Another phenomenon  that comes to mind in this respect is the marginal stability curves in N=2 SYM theory 
and other supersymmetric systems \cite{CMS}. When crossing these curves, quantum states may appear and disappear.
However, the mechanism for this is quite different there.  

Let us make somewhat unusual  conclusion listing again not the results obtained (that was done above), but rather
 the points which are not yet clear.

 \begin{enumerate}
 \item It is not clear whether the spectral problem (\ref{Schrgauge}) is well posed for the potential (\ref{2n+1}) 
and, if yes, what is its spectrum. Is the evolution operator unitary ?
 \item It is not clear whether one can formulate the spectral problems with discrete spectrum in the 
dashed region in Figs. 9,10 by
resolving the gauge constraint at the classical level with a clever gauge choice.
 \item
It is not clear why, in contrast to the complex oscillator case, we have not found  for the potential (\ref{2n+1}) 
a spectral problem involving only negative energy states
(the cryptoghosts !) and related to the sets of classical orbits with negative energies. Can such problem be formulated  ?

\end{enumerate}

The final remark is that crypto-Hermitian systems may prove to be something more than a formal mathematical exercise. They can bear
 relevance for physics. Our own  interest to these problems stems mainly from their relationship to  higher-derivative
systems. And we believe (the arguments were presented in Ref.\cite{TOE}) that the undiscovered yet fundamental 
Theory of Everything is a higher-derivative field theory ({\it not}  string theory) living in higher-dimensional space-time.
 
I am indebted to P. Dorey, L. Mezincescu, and D. Robert for useful discussions.

\end{document}